\documentclass[aps,pra,twocolumn,superscriptaddress,longbibliography,showpacs]{revtex4-1} 

\usepackage{amsmath}
\usepackage{graphicx,color,braket}
\usepackage{amsfonts}
\usepackage{amssymb}
\usepackage{mathrsfs}
\usepackage{multirow}
\usepackage{physics}
\usepackage[colorlinks = true]{hyperref}
\usepackage[T1]{fontenc}

\newcommand{\T}{\mathcal{T}}

\usepackage{color,braket}

\newcommand{\be}{\begin{equation}}
\newcommand{\ee}{\end{equation}}

\newcommand{\bla}{\color{black}}

\begin{document}

\title{Ruling out the class of statistical processes involving two noninteracting
identical particles in two modes}
\author{S. Aravinda}\email{sakshiprajne@gmail.com}
 \affiliation{Department of Physics, \\ Indian Institute of Technology Madras, Chennai 600036, India}
\begin{abstract}
In the framework of  Generalized  probabilistic theories (GPT),  we illustrate
a  class of   statistical  processes  in  case  of  two  non-interacting  identical
particles  in two  modes that  satisfies  a well  motivated notion  of
physicality       conditions   namely  the  double stochasticity and  the  no-interaction condition      proposed       by       Karczewski
et. al. (Phys. Rev. Lett. 120, 080401 (2018)), which  can not  be realized 
through  a  quantum mechanical process.  This class of statistical process is ruled
out by an additional requirement called  the  evolution condition imposed on two particle evolution.  We also show that any statistical 
process of two noninteracting identical particles in two modes that satisfies all  of the  three physicality conditions can be realized  within quantum mechanics using the beam  splitter operation. 

\end{abstract}

\maketitle

\section{Introduction}

Out of many features due to which quantum mechanics (QM) deviates from
classical  probability theory,  indistinguishable nature  of identical
particles, and  non-local nature \cite{EPR35} of  quantum correlations
remain   as   prominent   features.    The   statistical   nature   of
indistinguishable  particles plays  a central  role in  our every  day
understanding  starting  from the  molecules,  atoms,  solids to  many
astronomical events.  The non-local nature exhibited by quantum theory
via violation of Bell-type  inequality \cite{Bel64} has revolutionized
the fundamental understanding  of Nature, and also  has contributed to
development of quantum information theory and computation.

 GPT  framework  begins with  adopting  the  formalism of  operational
 quantum  mechanics which  was  advocated by  quantum logic  community
 \cite{mackey2004mathematical,    ludwig2012foundations,   BB93}    in
 pre-quantum information  era. Later,  due to  the advancement  in the
 quantum information  theory, GPT was formulated  using Euclidean real
 probabilistic vector spaces to  describe the state space \cite{Har01,
   Har03,  Man03}.  It  is  considered   as  one  of  the  fundamental
 frameworks  to study  the quantum  correlations in  any probabilistic
 theory  \cite{Har01,  Har03,  Man03,   BBL+07,  Bar07,  Spe05,  BW11,
   hardy2011foliable, CDP10}, and to  find the operationally motivated
 information   theoretic  axioms   for  quantum   theory\cite{  Dar07,
   chiribella2016quantum, chiribella2017quantum,MM11, DB11, BMU14}.

The quest  for the physical  axioms to  characterize QM began  with an
attempt  by  Popescu   and  Rohrlich  in  which   they  constructed  a
probabilistic  model  (later  called PR-box)  for  maximally  nonlocal
correlations   that  violate   CHSH   inequality   to  its   algebraic
maximum. The impossibility of realizing  PR-box correlations in QM was
due  to  Tsirelson  \cite{Cir80},  even though  the  correlations  are
non-local as  well as  non-signalling.  This  prompted the  search for
different physical principles to reproduce quantum mechanical bound in
case of CHSH inequality \cite{Cab13b,FSA+13,Paw+09,Nav10,Oppenheim10}.

GPT framework, which can accommodate any probabilistic physical theory
facilitated such  constructions, thus provided a  deeper understanding
of the nature  of QM, and in  tern that of Nature.  Motivated by these
developments, GPT  framework has been extended  to accommodate general
relativity \cite{Har16} and  indefinite causal structure \cite{Har09}.
 Hardy has initiated the program to extend GPT framework to field
theories,  and  has  provided  a manifesto  for  constructing  quantum
gravity \cite{Har18}.

Recently,  Karczewski  et.   al.   \cite{Kar18},  proposed  a  general
probabilistic  framework   to  deal  with   non-interacting  identical
particles.   In this  work, the  authors formulated  a well  motivated
physical   principle   called   consistency   condition   similar   to
no-signaling condition that any  theory with non-interacting identical
particles should  satisfy, and  provided an  example of  a statistical
process with three  identical particles in three  mode, that satisfies
consistency  condition but  cannot be  realized in  QM. Following  the
framework of  Karczewski et.  al. \cite{Kar18}, in  this work  we show
that their exists much simpler configuration, i.,e two non-interacting
identical particles in two mode which satisfies consistency condition,
yet fails to  produce such process in  QM.  We also show  that an extra
physicality condition proposed by Karczewski et. al., which recovers quantum
mechanical statistics in case of three particles in three modes, 
can  also be  used  to rule  out proposed  impossible  process in  two
identical particles in two mode case. We call this principle `the evolution principle', 
and identify that the class of physical theories involving generalized statistics 
of identical particles, interpolated between fermions and bosons,  \cite{Gren91,Fiv90,Gren92} also satisfy
this principle.  We show that any process which satisfies  all  of the 
three physicality conditions can be realized by a quantum mechanical beam splitter configuration. 

\section{Generalized probability theory framework}
The  basic  ingredients  of  GPT's  are  states,  transformations  and
measurements,  and  are  formulated  in the  language  independent  of
Hilbert space formalism  so that it can  accommodate any probabilistic
physical theory.  In case of distinguishable  particles, the framework
is very  well developed, and has  been used as a  cornerstone for both
foundational understanding  as well  as for  applications. In  case of
foundations,  the  framework  is   used  to  provide  the  physicality
conditions              that             characterizes              QM
\cite{Cab13b,FSA+13,Paw+09,Nav10,Oppenheim10}  ,   to  understand  the
limits  and advantages  of general  correlations \cite{Bar07,  BLM+05,
  BBL+06,  BM06,  BBL+07,  Bar03,  JH14}, and  to  find  an  axiomatic
formulation  of QM  \cite{Har01,CDP11, hardy2011reformulating,  MM11,
  masanes2013existence}. On  the application  side, it has  provided a
methodology \cite{GRINBAUM201722} of device independence certification
of many quantum information theoretic and quantum computational tasks.

Recently Karczewski  et al,\cite{Kar18} developed a  GPT framework for
noninteracting identical particles, which we briefly review here.

Consider $N$ particles in $M$ modes.  The state of identical particles
are determined  by particle  occupation number in  each mode,  $\phi =
\{n_1, n_2, \cdots n_M\}$, where $n_1, n_2, \cdots n_M$ are occupation
number in mode $\{1,2, \cdots M\}$  with $\sum_{i=1}^{M} n_i = N$. The
state probability vector $\Phi$ is a d-dimensional vector representing
the probability  distribution of  particles over  all modes.   The GPT
framework  consists of  an initial  state $\Phi_i$,  an evolution  (or
transformation) $\T$ given  by a stochastic matrix, and  a final state
$\Phi_f$.

For example, consider a single boson in two mode with a symmetric beam
splitter (BS) transformation.  The set of occupation number states are
$\{1,0\}$  and  $\{0,1\}$.   The  state  probability  vector  $\Phi  =
(P(\{1,0\}),  P(\{0,1\}))^T$, where  $P(\{1,0\})$  is the  probability
distribution of particles.  The transformation matrix $\T^{(1)}_{BS}$,
for a BS for single particle in two modes is given by

\begin{equation}
 \T^{(1)}_{BS} = \frac{1}{2}\left(\begin{array}{cc}
                        1 & 1  \\
                        1 & 1 \end{array} \right).
                        \label{eq:BS1}
                     \end{equation}

Similarly, a GPT framework for two  bosons in symmetric BS is given by
representing the state $\Phi = (P(\{2,0\}), P(\{1,1\}),P(\{0,2\}))^T $
and the transformation $ \T^{(2)}_{BS}$ by
\begin{equation}
 \T^{(2)}_{BS} = \left(\begin{array}{ccc}
                        1/4 & 1/2 & 1/4  \\
                        1/2 & 0 & 1/4 \\
                        1/4 & 1/2 & 1/4 \end{array} \right).
                        \label{eq:BS2}
                     \end{equation}

\section{Physicality conditions}
 
In case of characterizing nonlocal correlations in Bell-type scenario,
the natural physicality condition is to satisfy relativistic causality
i.e, impossibility  of instantaneous communication.   The no-signaling
condition  states that  the marginal  probability distribution  of one
party  should not  be affected  by the  choice of  observables by  any
another spatially separated party. Similarly, in case of contextuality
\cite{KS67} it  is Gleason's no-signaling (also  called no-disturbance
in literature) which acts as the physicality or consistency condition.

The consistency condition for  non-interacting identical particles has
to  consider  the  non-interacting  character.   This  requirement  as
formulated in Ref. (\cite{Kar18}),   which  we  call  here
no-interaction criteria.
\begin{description}
 \item[No-interaction]  The   transformation  of  a   single  particle
   distribution should  not be affected  by the presence of  any other
   particle.
\end{description}

The   precise  mathematical   characterization  requires   defining  a
transition  matrix  $\mathcal{R}^{(N)}$ with  elements  $R^{(N)}_{ij}$
which specify the  transition of $N$-particle state  $\Phi_j$ to $N-1$
particle  state  $\Phi_i$,  by  randomly removing  one  particle  from
$N$-particle state:
\begin{equation}
 \Phi^{(N-1)} = \mathcal{R}^{(N)} \Phi^{(N)}. 
 \label{eq:rem}
\end{equation}

Any  $K$-particle state  can be  obtained from  $N$-particle state  by
sequentially removing single particle.  For example, a transition from
$3$  particle  state to  single  particle  state  can be  obtained  as
$\Phi^{(1)}  =  \mathcal{R}^{(2)}\mathcal{R}^{(3)}\Phi^{(3)}$.   Thus  a
transition  matrix $\mathcal{R}^{(N\rightarrow  K)}$ for  $N$-particle
state to $K$-particle state is given by,
\begin{equation}
 \mathcal{R}^{(N\rightarrow K)} = \mathcal{R}^{(K+1)} \cdots \mathcal{R}^{(N-1)}\mathcal{R}^{(N)}.
\end{equation}

With   these  mathematical   devices,  the   no-interaction  condition
constrain the allowed transformations $\T$ as,
\begin{equation}
 \mathcal{R}^{(N\rightarrow K)} \T^{(N)} \Phi_i = \T^{(K)} \mathcal{R}^{(N\rightarrow K)} \Phi_i , \forall i .
 \label{eq:ni}
\end{equation}
This  means  that the  state  probability  vector obtained  by  first
reducing  an  $N$-particle  state  to $K$-  particle  state  and  then
transferring a $K$- particle state  must be same as first transferring
an $N$-particle state and then reducing it to a $K$-particle state.

This will  be evident by  considering an elementary system with  N=2 and M=2,
for which $\mathcal{R}^{(2)}$ is given by
\begin{equation}
 \mathcal{R}^{(2)} = \frac{1}{2} \left( \begin{array}{ccc}
                      2 & 1 & 0 \\
                      0 & 1 & 2 \end{array} \right).
                      \label{eq:R2}                   
                      \end{equation}
The no-interaction condition constraints any  $\T^{(2)}$ that satisfies, 
\begin{equation}
 \mathcal{R}^{(2)} \T^{(2)} \Phi_i^{(2)} =   \T^{(1)} \mathcal{R}^{(2)} \Phi_i^{(2)},
 \label{eq:R2ni}
\end{equation}
for all states $\Phi_i$. It is  very clear from Eq. (\ref{eq:BS1}) and
Eq.   (\ref{eq:R2}), that  the  symmetric BS  transformation given  in
Eq. (\ref{eq:BS2}) satisfies no-interaction condition (\ref{eq:R2ni}).

No increase of entropy after
transformation demands {\it double stochasticity} condition which is stated as follows:

\begin{description}
 \item[Double stochasticity] The transformation matrix $\mathcal{T}$ must be  doubly stochastic. 
\end{description}

The final condition on the transformation $\mathcal{T}$ to be physical is the 
{\it evolution principle} which is satisfied within QM \cite{TA13,Lou98}, and it can be stated as follows: 
\begin{description}
 \item[Evolution principle] The evolution of the states in
which  all the  particles are  in  same mode  should be  equal to  the
evolution generated by its single particle counterpart.
\end{description}

In the two particle case, this principle is written as:  
\begin{equation}
 \T^{(2)} \Phi^{(2)} = \T^{(1)} \Phi^{(1)} \times \T^{(1)} \Phi^{(1)} . 
 \label{eq:imp}
\end{equation}

\section{Characterization of two non-interacting identical particles}
 In this section we characterize  the transformation of two non-interacting 
identical particle in two modes satisfying the conditions of double stochasticity, no-interaction and the evolution principle. \bla 
It is shown that for two non-interacting  identical particles in two modes,  any transformation that satisfies all the 
three physicality conditions can be realized by quantum mechanical identical particles in a general beam splitter transformation. 

\subsection{GPT characterization \label{sec:gptchar}}
Consider a general transformation $\mathcal{T}_{g}^{(2)}$ of two identical particles in two modes that satisfies doubly stochastic condition:

\begin{equation}
 \T^{(2)}_{g} = \left(\begin{array}{ccc}
                        \alpha_1 & \alpha_2 & 1-\alpha_1 -\alpha_2  \\
                        \alpha_3 & \alpha_4 & 1-\alpha_3 -\alpha_4 \\
                        1-\alpha_1 - \alpha_3 & 1-\alpha_2 -\alpha_4 & -1+\sum_{i=1}^4 \alpha_i  \end{array} \right).
                        \label{eq:BSg}
\end{equation}
Similarly, the general transformation $\mathcal{T}_g^{(1)}$ of a single particle in two modes is:

 \begin{equation}
 \T^{(1)}_{g} = \left(\begin{array}{cc}
                        \beta & 1-\beta  \\
                        1-\beta & \beta \end{array} \right).
                        \label{eq:BS1g}
                     \end{equation}
Now, applying the no-interaction condition (\ref{eq:R2ni}) on the general transformations (\ref{eq:BSg}) and (\ref{eq:BS1g}), 
constrains the values of $\alpha_i$ as 
\begin{align}
 \alpha_4 &= 1-2\alpha_2 \nonumber \\
 \alpha_3 &= 2(\beta - \alpha_1). 
 \label{eq:nicon}
\end{align}
After the application of no-interaction condition,  the general transformation is a three parameter family which can be written as

\begin{equation}
 \T^{\prime (2)}_{g} = \left(\begin{array}{ccc}
                        \alpha_1 & \alpha_2 & 1-\alpha_1 -\alpha_2  \\
                        2(\beta-\alpha_1) & 1-2 \alpha_2 & 2(-\beta +\alpha_1 +\alpha_2 \\
                        1+\alpha_1 - 2\beta & \alpha_2 & 2\beta - \alpha_1 -\alpha_2  \end{array} \right). 
                        \label{eq:BS2g}
\end{equation}

Further application of evolution principle  (\ref{eq:imp}) to Eq. (\ref{eq:BS2g})  relates the two parameters $\alpha_1$ and $\alpha_2$ by the equations 
\begin{align}
 \alpha_1 & = \beta^2 \nonumber \\
 \alpha_2 & = 2(\beta - \beta^2 ).
 \label{eq:fina}
\end{align}
 The general transformation $\T^{(2)}_g$ which satisfy all the three physicality conditions
is a single parameter family represented by parameter $\beta$ and it is given as 

\begin{equation}
	\T^{\prime (2)}_{g} (\beta) = \left(\begin{array}{ccc}
		\beta^2 & 2\beta (1-\beta) & (1-\beta)^2  \\
        2\beta (1-\beta) & 1 -4\beta (1-\beta) & 2\beta (1-\beta) \\
        (1-\beta)^2 & 2\beta (1-\beta) & \beta^2  \end{array} \right). 
                        \label{eq:BS2gfin}
\end{equation}

\subsection{Quantum mechanical characterization}
The input-output relation for boson annihilation operators for general  two mode BS can be 
written as 

\begin{align}
 \left( \begin{array}{c}
 \hat{b_1} \\
 \hat{b_2}\\
       \end{array} \right) & = U_{BS} \left( \begin{array}{c}
 \hat{a_1} \\
 \hat{a_2}\\
       \end{array} \right) \nonumber \\
       & = \left(\begin{array}{c c }
                  u_{11} & u_{12} \\
                  u_{21} & u_{22}
                 \end{array}\right)
 \left( \begin{array}{c}
 \hat{a_1} \\
 \hat{a_2}\\
       \end{array} \right),
       \label{eq:gBS}
       \end{align}

where $(\hat{a_1}, \hat{a_2})$ and $(\hat{b_1}, \hat{b_2})$ are the input-output boson annhilation 
operators. The output bosonic commutation relation $[\hat{b}_i,\hat{b}_j] = \delta_{ij}$ restricts the elements of  the transformation $U$ with $|u_{11}|^2 + |u_{12}|^2 = 1$, $ |u_{21}|^2 + |u_{22}|^2 = 1$ and $u_{11}u_{11}^\ast + u_{22}u_{22}^\ast = 0$. These restrictions imply that the BS transformation has to be an unimodular representation of subgroup $SU(2)$ \cite{CampBS}, and the simplest representation of $SU(2)$ that realizes the BS transformation is 
\begin{equation}
 U_{BS} = \left( \begin{array}{c c}
             \cos{\theta} & \sin{\theta}  \\
             -\sin{\theta} & \cos{\theta}
            \end{array}\right).
\label{eq:Ubs}
\end{equation}

The corresponding two particle $\mathcal{T}^{(2)}_{QM}$ and single particle $\mathcal{T}^{(1)}_{QM}$  transformations  in the GPT  can be written as
\begin{equation}
 \mathcal{T}^{(1)}_{QM} = \left( \begin{array}{c c}
                                  \sin^2{\theta} & \cos^2{\theta} \\
                                  \cos^2{\theta} & \sin^2{\theta}
                                 \end{array}\right)
                                 \label{eq:BS1qm}
\end{equation}
and
\begin{widetext}
\begin{equation}
 \mathcal{T}^{(2)}_{QM} = \left( \begin{array}{c c c}
                                  \sin^4{\theta} & 2 \sin^2{\theta}\cos^2{\theta} & \cos^4{\theta} \\
                                  2 \sin^2{\theta}\cos^2{\theta} & \cos^2{2\theta} & 2 \sin^2{\theta}\cos^2{\theta} \\
                                  \cos^4{\theta} & 2 \sin^2{\theta}\cos^2{\theta} & \sin^4{\theta}
                                 \end{array} \right).
                                 \label{eq:BS2qm}
\end{equation}
\end{widetext}

 The constraint on the general transformation of two noninteracting identical particle in two modes  that satisfies all the three conditions of physicality is characterized in Section (\ref{sec:gptchar}) and the allowed single parameter general transformation is given in Eq. (\ref{eq:BS2gfin}). Any transformation that satisfies all the three physicality conditions can be realized in quantum mechanical way by a BS transformation (\ref{eq:BS2qm}) by substituting $\beta = \sin^2{\theta}$
in  Eq.(\ref{eq:BS2gfin}). It is important to note that the given QM realization is not unique. The general 
BS transformation can be represented by including the phase factors. There can be many BS unitaries that can realize the given GPT transformation. 
\bla

\section{Impossible process}
 Karczewski  et.al.,\cite{Kar18}  provided  an example  of  a  process
 involving $N=3$ in $M=3$ system with a transformation $\T^{(3)}$ that
 exceeds the bunching  probability bound given in QM,  even though the
 transformation   $\T^{(3)}$    satisfies   no-interaction   condition
 (\ref{eq:ni}), as  well the matrix representing  $\T^{(3)}$ is doubly
 stochastic.   The  given  transformation   (a.k.a.  process)  is  not
 realizable in QM, similar to  maximally non-local PR-box which cannot
 be  realized in  QM.    Here we report a class of transformations $\T^{(2)}$ which satisfies 
 double stoachasticity and no-interaction conditions, that can not be realized in QM. The class of transformations 
 $\T^{\prime (2)}$ in Eq. (\ref{eq:BS2g}) that violate the evolution condition given in Eq. (\ref{eq:fina}) can not be realized within QM. 
 
An elementary
 example  involving $N=2$  in $M=2$,  which satisfies  no-interaction
 condition  (\ref{eq:R2ni})  and  doubly  stochastic  but  no  quantum
 mechanical unitary transformation that can realize such a process, can be given by the 
 transformation $\T_{(imp)}^{(2)}$ 
 \begin{equation}
  \T_{(imp)}^{(2)} = \left( \begin{array}{ccc}
                             1/2 & 0 & 1/2 \\
                             0   & 1 & 0   \\
                             1/2 & 0 & 1/2 
                            \end{array} \right).
                            \label{eq:timp}
\end{equation}
It can be easily verified that $\T_{(imp)}^{(2)}$ is doubly  stochastic, and 
satisfies no-interaction condition (\ref{eq:R2ni}) with respect to the $\T^{(1)}_{BS}$ in Eq. (\ref{eq:BS1}). 
We can use evolution  principle to rule out  $\T_{(imp)}^{(2)}$, as the
transformation $\T_{(imp)}^{(2)}$ violates   it. This  can be  seen by
noting    that    $P^{(20)}_{(20)}   \neq    P^{(10)}_{(10)}    \times
P^{(10)}_{(10)}$.

\section{Evolution condition as physical condition}
The notion of physicality condition is obtained by noting the general principle satisfied by the required physical theory. For example, the most general physicality conditions like  the  no-signaling  and  the  laws of thermodynamics arise  from the fact that all known physical theories  satisfy  these conditions. The quest for deriving QM from well formulated physical principles  began with a  more preliminary task of bounding the nonlocal correlations to Tsirelson's value (bound) \cite{Cir80}. In the process, many physicality conditions  were  formulated,  which bound nonlocal  correlations  to quantum value \cite{Cab13b,FSA+13,Paw+09,Nav10,Oppenheim10}.  The basic idea is that these principles lead to the predictions satisfied by classical and quantum mechanical theories. On the contrary, violating any of these principles lead to predictions that are not in consonance with that of either classical or quantum theories.

Similarly, as noted in the ref.\cite{Kar18}, and was shown in ref \cite{Lou98}, the evolution principle is  satisfied by classical and quantum theories making it  a viable  candidate of physical principle.   If the proposed physicality principle  is satisfied  by broad range of physical theories, then the principle will have the broad  range of  theoretical validity. As a first step in this direction, we show that  in a  two mode beam splitter scenerio,  the evolution principle is satisfied by more generalized statistics of identical particles interpolated between fermions and bosons generated by deformed Fermi and Bose algebras \cite{Gren91,Fiv90,Gren92}.  The extended algebra are called q-deformed algebra and the excitations are called \textit{quons}. The operator algebraic relation between creation ($\hat{a}^\dagger$) and annhilation ($\hat{a}$) operators are given as 
\begin{equation}
 \hat{a}_i\hat{a}^\dagger_j - q \hat{a}^\dagger_j\hat{a}_i = \delta_{ij} I
 \label{eq:quanal}
\end{equation}
where $-1\leq q \leq 1$ is the deformation factor. This algebraic relations leads to the following relations: 
\begin{align}
 \hat{a}_i \ket{0} = 0  \quad  & \quad \hat{a}^\dagger_i \hat{a}_i \ket{n_i} = v_n \ket{n_j} \nonumber \\
 \hat{a}^\dagger_i \ket{n_i} = \sqrt{v_{n+1}} \ket{(n+1)_i} & \quad \hat{a}_i \ket{n_i} = \sqrt{v_n} \ket{(n-1)_i} \nonumber \\
 (\hat{a}^\dagger_i)^n \ket{0} = \sqrt{v_n v_{n-1} \cdots v_1} \ket{n_i} &
\end{align}
where 
\begin{equation}
 v_n = \begin{cases}
        \sum_{m=0}^{n-1} v^m = 1+ q v_{n-1} & , n \geq 2 \\
        1 &, n = 1.
       \end{cases}
\end{equation}

Consider the linear transformation as BS operation given in Eq. (\ref{eq:gBS}), where 
$(\hat{a}_1, \hat{a}_2)$ and $(\hat{b}_1, \hat{b}_2)$ are the input-output quon annhilation 
operators. Denote $\sqrt{T} = u_{11} = u_{22}$ and $\sqrt{R} = u_{12} = - u_{21}$ as reflection and 
transmission co-efficients of BS with $R + T = 1$. 

The single quon in two mode  transforms  according to  the  Eq. (\ref{eq:gBS})  as  :
\begin{align}
 \ket{\phi_{in}} & = \hat{a}^\dagger_1 \ket{00} \nonumber \\
				 & \rightarrow u_{11} \ket{1,0} + u_{21} \ket{0,1}.
\end{align}
Accordingly, the single quon statistics are  :  $P^{(1,0)}_{(1,0)} = T,  P^{(1,0)}_{(0,1)} = R,  P^{(0,1)}_{(1,0)} = R, P^{(0,1)}_{(0,1)} = T$. Beginning with single quon in both the modes, we get 
\begin{widetext}
\begin{align}
 \ket{\phi_{in}} & = \hat{a}^\dagger_1 \hat{a}^\dagger_2\ket{00} \nonumber \\
                 & \rightarrow (u_{11}u_{12} (\hat{b}^\dagger_1)^2 + u_{21}u_{22}(\hat{b}^\dagger_1)^2 + u_{11}u_{12}\hat{b}^\dagger_1\hat{b}^\dagger_2 + u_{21}u_{12}\hat{b}^\dagger_2\hat{b}^\dagger_1) \ket{00}. 
\end{align}
\end{widetext}
From this we get $P^{(1,1)}_{(2,0)} = P^{(1,1)}_{(0,2)} = RT (1+q)$ and  $P^{(1,1)}_{(2,0)} = R^2 + T^2 -2q RT $. Similarly we get $P^{(2,0)}_{(2,0)} = P^{(0,2)}_{(0,2)} = T^2$ ,  $P^{(2,0)}_{(0,2)} = P^{(0,2)}_{(2,0)} = R^2$ and 
$P^{(2,0)}_{(1,1)} = P^{(0,2)}_{(1,1)} = 2RT$. From this it is easy to see that the quon statistics satisfies evolution principle (\ref{eq:imp}) for two quons in two modes. Like for example, note  $P^{(2,0)}_{(2,0)} = P^{(1,0)}_{(1,0)} \times P^{(1,0)}_{(1,0)} = T^2$.

\bla
\section{Conclusion}
Non-local  nature  of  quantum  correlations is  one  of  the  salient
features  that   dramatically  deviates  from  our   understanding  of
classical correlations  which purely  originate from  human ignorance.
Maximal  nonlocality  beyond QM  exhibited  by  PR-box, has  laid  the
foundation  for   exploring  the   possibility  of   finding  physical
principles behind many properties of QM, and provided a methodological
application  in  terms  of   device  independent  characterization  of
information theoretic tasks.

Indistinguishable nature  of quantum identical particles  deviates our
world  view  of  undertanding  Nature from  a  classical  description.
Recent computational advantages in  case of Boson sampling \cite{AA11}
have  solely emerged  from the  indistinguishable nature  of identical
quantum particles.   Formulating and  studying identical  particles in
GPT, and  identification of  more general  principles that  single out
quantum mechanical statistics is important.

 In this work,  we consider an elementary setting  of two noninteracting
 identical particles in two modes, in which we provide a class of transformations  that   satisfies well defined notion
 of physicality conditions, yet cannot  be realized in QM. A principle
 proposed by  Karczewski et.  al.,\cite{Kar18}, where  they considered
 three particles in three modes, has been applied in this work for the
 case of  a more elementary system  of two identical particles  in two
 mode to rule out such an  impossible process. This provides a greater
 insight  to  identify  physicality  condition which  will  provide  a
 necessary and sufficient condition for bounding the quantum mechanical
 probabilities  in case  of  many identical  particles  in multi  mode
 settings.

 \begin{acknowledgments}
  I  would like  to thank  A. Mukherjee,  S. Ghosh,  R. Srikanth,  for
  helpful discussions,  and S. Utagi  for carefully reading  the draft
  and suggesting modifications.
 \end{acknowledgments}

\bibliography{thesis,os,preeti,research}
\end{document}